\documentclass[final,5p,times,twocolumn]{elsarticle}

\usepackage{graphicx,amsmath,amssymb,bm,multirow}
\usepackage{amsthm}
\usepackage{amsfonts}

\usepackage[usenames,dvipsnames]{color}

\usepackage[nodots]{numcompress}

\begin{document}

\begin{frontmatter}

\title{Enhanced collectivity in neutron-deficient Sn isotopes in energy functional based collective Hamiltonian}

\author{Z. P. Li\fnref{1}}
\author{C. Y. Li\fnref{1}}
\author{J. Xiang\fnref{1}}
\author{J. M. Yao\fnref{1,2}}
\author{J. Meng\fnref{3,4,5}}

\address[1]{School of Physical Science and Technology,
             Southwest University, Chongqing 400715, China}
\address[2]{Physique Nucl\'eaire Th\'eorique, Universit\'e Libre de Bruxelles, C.P. 229, B-1050 Bruxelles, Belgium}
\address[3]{State Key Laboratory of Nuclear Physics and Technology, School of Physics,
Peking University, Beijing 100871, China}
\address[4]{School of Physics and Nuclear Energy Engineering, Beihang University, Beijing 100191, China}
\address[5]{Department of Physics, University of Stellenbosch, Stellenbosch, South Africa}

%
%

\begin{abstract}
The low-lying collective states in Sn isotopes are studied by a
five-dimensional collective Hamiltonian with parameters determined
from the triaxial relativistic mean-field calculations using the
PC-PK1 energy density functional. The systematics for both the
excitation energies of $2^+_1$ states and $B(E2;0^+_1\to 2^+_1)$
values are reproduced rather well, in particular, the enhanced $E2$ transitions
in the neutron-deficient Sn isotopes with $N<66$.  We show that the gradual
degeneracy of neutron levels $1g_{7/2}$ and $2d_{5/2}$
around the Fermi surface leads to the increase of level density
and consequently the enhanced paring correlations from $N=66$ to $58$.
It provokes a large quadrupole shape fluctuation around the
spherical shape, and leads to an enhanced collectivity in the
isotopes around $N=58$.
\end{abstract}

\begin{keyword}
covariant energy density functional \sep
collective Hamiltonian \sep
low-lying states \sep
electromagnetic transition \sep
Sn isotopes

\PACS 21.60.Jz \sep 21.60.Ev \sep 21.10.Re \sep 21.10.Tg
\end{keyword}

\end{frontmatter}

%
%
The evolution of shell structure and nuclear collectivity along
either isotopic or isotonic chains, inferred from the systematics of
the nuclear deformation energy surface~\cite{Meng.05}, the nucleon separation
energy~\cite{Casten07,prc1104Liang}, the excitation energy of low-lying
state and the electric quadrupole transition
strength~\cite{Cejnar10,Li.09a}, becomes of great interest in nuclear physics. In
recent years, a number of experiments have measured the $B(E2;
0^+_1\to 2^+_1)$ transition strengths in neutron-deficient Sn
isotopes with
$N<66$~\cite{Banu.05,Ced.07,Vam.07,Orce.07,Eks.08,Doo.08,Kumar.10}.
The deduced $B(E2)$ value increases when going from $N=66$ to
$N=64$, and then remains roughly constant within the experimental
uncertainties when decreasing the neutron number down to $N=56$.
This picture deviates from the parabolic trend, with a peak at
mid-shell, predicted by the single $j$-shell seniority
model~\cite{Talmi.93}. Furthermore, the $B(E2)$ values deduced from
the measurement of the lifetimes of the first excited $2^+$ states
in $^{112,114,116}$Sn at the UNILAC accelerator of the Gesellschaft
f\"ur Schwerionenforschung (GSI) are overall smaller than the
previously reported values and present a shallow minimum at $N =
66$~\cite{Jung.11}. These findings have generated a renewal interest
in the study of low-lying states in Sn isotopes.

The theoretical description of the enhanced $B(E2; 0^+_1\to 2^+_1)$
transition strength in the neutron-deficient Sn isotopes is not
straightforward. Similarly as the seniority scheme, the seniority
truncated large-scale shell model calculations predicted a nearly
parabolic behavior of $B(E2)$ values with a maximum at
$N=66$~\cite{Banu.05}. By using new interactions and including core
polarization terms up to both the third order and $5\hbar\omega$
core excitations in the shell model calculations, the predicted
$B(E2)$ values exhibit a shallow maximum around $N= 68-70$, but are
still far away from the experimental data \cite{Eks.08}.
Shell model calculations in the $sdgh$ major shell using CD-Bonn and Nijmegen1 two-body effective
nucleon-nucleon (NN) interactions were also carried out  for the mid-heavy
even-even Sn isotopes in Refs.~\cite{Dikmen09,Guazzoni11}.
An overall good agreement with the observed energy spectra
of Sn isotopes is found. However, the $E2$ transitions are not presented.
Recently, a quasiparticle-phonon model (QPM)
was adopted to investigate the evolution of the $2^+_1$ state in Sn
isotopes~\cite{Iud.11}. The calculation reproduced the trend of the
energies and, partly, the observed deviations of the $E2$ strengths
from the parabolic behavior. It was shown that such an asymmetric
trend was the consequence of several factors: single-particle
energies, polarization of the $N = Z = 50$ core, interplay between
pairing plus quadrupole, and quadrupole pairing
interactions~\cite{Iud.11}, all of which, however, are quite
model-dependent.

In the framework of nuclear energy density functional theory or
self-consistent Hartree-Fock-Bogoliubov (HFB) approach, due to the
robust $Z=50$ shell gap, the ground states of all the Sn isotopes
are dominated by the spherical configuration. For the $2^+_1$
states, which are expected to be a one-phonon state caused by
surface vibration around the spherical shape, the theoretical
studies are obviously beyond the mean-field approximation. For this
purpose, one can either introduce random-phase approximation (RPA)
or generator coordinate method (GCM), where the polarization of $N=Z=50$ core
in the shell model can be taken into account by particle-hole excitation in RPA
or configuration mixing of different shapes in GCM.
The implementations of the
quasiparticle RPA on top of the nonrelativistic and relativistic
mean-field calculations were carried out to investigate the
evolution of low-lying states along the Sn isotopic
chain~\cite{Ter.04,Ans.05,Ans.06,Tian09}. In Refs.~\cite{Ans.05,Ans.06},
the systematics of $B(E2; 0^+_1\to 2^+_1)$ values can be roughly
reproduced. However, these studies were focused on the substantial
rise of the $B(E2)$ values at neutron magic numbers 50 and 82, and
did not provide any interpretation for the enhanced collectivity in
the neutron-deficient Sn isotopes.

In recent years, the GCM has been introduced into the
self-consistent mean-field approaches by combing with projection
techniques in the modern energy density functional
calculations~\cite{Valor00,Guzman02,Niksic06,Bender08,Yao08,Yao09,Yao10,Rodriguez10}.
The dynamic correlation effects related to the symmetry restoration
and quadrupole fluctuation (along both $\beta$ and $\gamma$ directions)
around the mean-field minimum could be taken into
account. These methods restricted to axial case have
been applied to study the low-lying collective excitation
spectra of the neutron-deficient Pb isotopes~\cite{Bender04,Guzman04}.
However, the application of these methods with triaxial
degree-of-freedom for systematic study is still much time-consuming.
Up to now, such kind of study is mostly restricted to light
nuclei~\cite{Yao11-Mg,Yao11-C} and some specific medium heavy
nuclei~\cite{Tomas11PRC,Tomas11PLB}.  As the Gaussian overlap
approximation of GCM, collective Hamiltonian with parameters
determined by self-consistent mean-field calculations is much simple
in numerical calculations, and has achieved great success in
description of nuclear low-lying states for deformed and
transitional
nuclei~\cite{Libert99,Prochniak04,Nik.09,Hinohara10,Li.09a,Li.09b,Li.10,Li.11,Niksic11,Yao11-lambda,Mei12,Del10}.

In the past decades, the covariant density functional theory (CDFT)
has achieved great success in the description of nuclear structure
over almost the whole nuclide chart, from light systems to
superheavy nuclei, and from the valley of $\beta$ stability to the
particle drip lines \cite{BHR.03,VALR.05,Meng.06}. Recently, the
implementation of the five-dimensional collective Hamiltonian (5DCH)
based on the CDFT has been developed and given successful
microscopic description for the nuclear shape transitions along both
the isotopic and isotonic
chains~\cite{Li.09a,Nik.09,Li.09b,Li.10,Li.11}.
In this work, we are going to apply this approach to study the low-lying collective
states, for the first time, in the semi-magic Sn isotopes.
Our aim is to provide a  microscopic mechanism for
the enhanced $E2$ transitions in the neutron-deficient Sn isotopes.

In the calculation, we use a recent parameterized
relativistic functional PC-PK1~\cite{Zhao.10}  for the particle-hole
channel, and a separable force~\cite{TMR.09a,NRV.10} for the
particle-particle channel. The solution of the equation of motion
for the nucleons is accomplished by an expansion of the Dirac
spinors in a set of three-dimensional harmonic oscillator basis
functions in Cartesian coordinates with 12 major shells. More
details about the calculations can be found in
Ref.~\cite{Yao09,Nik.09,Xiang.12}. The intrinsic triaxially deformed
states are obtained as solutions of the self-consistent RMF+BCS
equations constrained on the mass quadrupole moments related to the
Bohr parameters $(\beta,\gamma)$ varying $\beta\in[0.0, 0.6]$ and
$\gamma\in[0^\circ, 60^\circ]$ with step size $\Delta\beta=0.05$ and
$\Delta\gamma=10^\circ$.

The nuclear collective excitations associated with three-dimensional
rotation and quadrupole fluctuations are described with a
five-dimensional collective Hamiltonian (5DCH). The dynamics of the
5DCH is governed by the seven functions of the intrinsic
deformations $\beta$ and $\gamma$: the collective potential $V_{\rm
coll}$, the three mass parameters: $B_{\beta\beta}$,
$B_{\beta\gamma}$, $B_{\gamma\gamma}$, and the three moments of
inertia $\mathcal{I}_k$. These functions are determined using
cranking approximation formula based on the intrinsic triaxially
deformed states. The diagonalization of the Hamiltonian yields the
excitation energies and collective wave functions that are used to
calculate observables~\cite{Nik.09}.

\begin{figure}[htb!]
\centering
\includegraphics[scale=0.26]{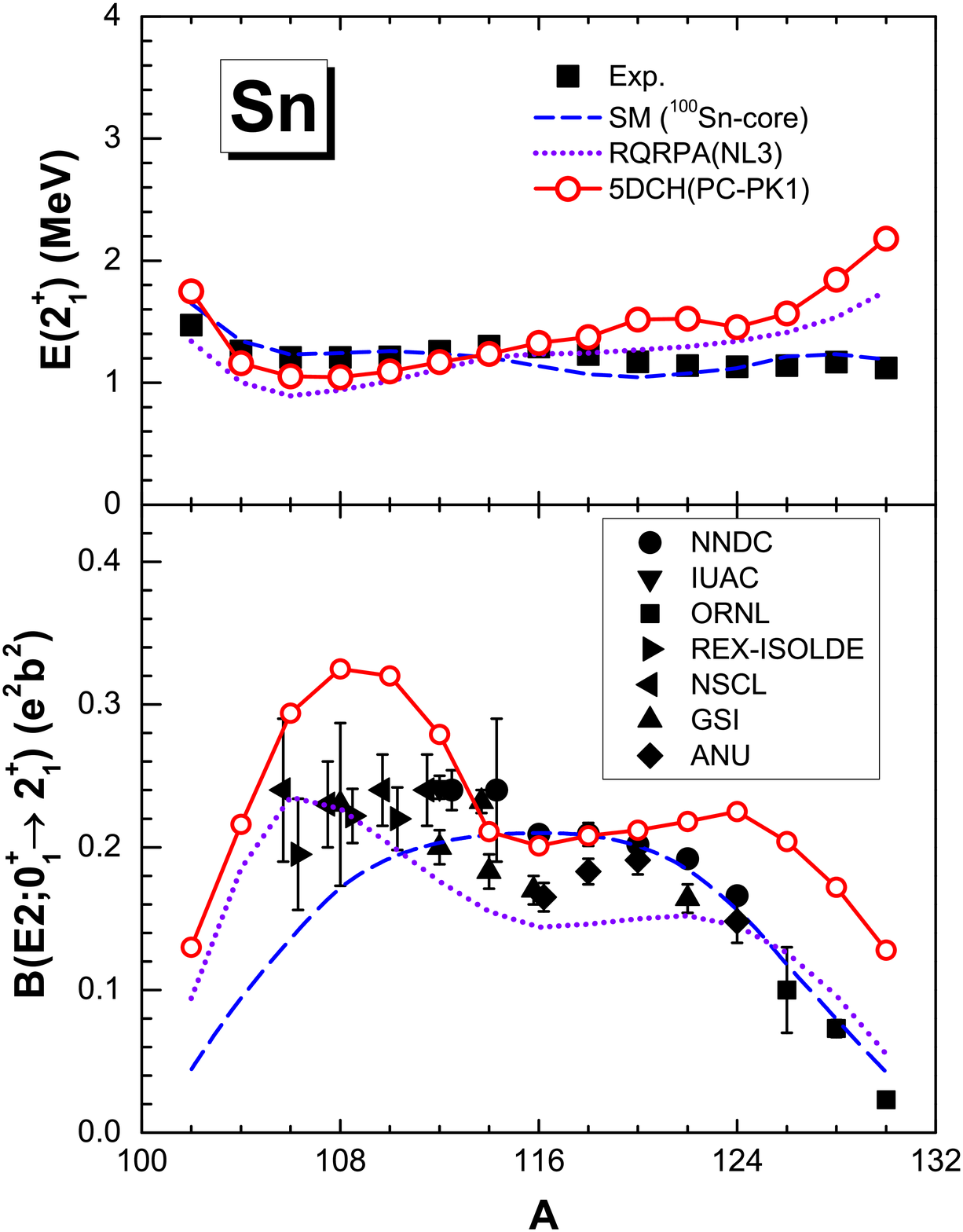}
\caption{\label{fig:BE2}(Color online)  Calculated $E(2^+_1)$ (upper
panel) and $B(E2;0^+_1\to 2^+_1)$ (lower panel) in Sn isotopic chain
by the shell model with $^{100}$Sn core (dashed line), RQRPA (dotted
line), and 5DCH based on PC-PK1 density functional (solid line with
open circles), in comparison with the data (solid symbols). In the
lower panel, the measured $B(E2;0^+_1\to 2^+_1)$ values from
different experiments are indicated with different symbols.}
\end{figure}

The upper panel of Fig.~\ref{fig:BE2} displays the excitation energies of $2^+_1$ states for Sn isotopes from our 5DCH calculations, in comparison with those by the shell model with $^{100}$Sn core and RQRPA calculations, as well as the data.
The trend of the excitation energies of $2^+_1$ states along the isotopic chain is reproduced quite well by the three models.
It is notable that our 5DCH calculations give quite similar results as those by the RQRPA calculations, although the realization of these two models for excitation states are quite different. Compared with the shell model calculation, the discrepancy between our 5DCH result and the data is relatively larger for $^{120-130}$Sn, from $\sim200$~keV to $\sim1$~MeV. However, it is worthwhile to note that the 5DCH is fully self-consistent and without any free parameter.
In addition, we find that, beside $2^+_1$ state, the excitation energies of other low-lying states for $^{108-116}$Sn are described as well as that by the recent shell model calculations~\cite{Dikmen09,Guazzoni11}. Taking $^{116}$Sn as an example one finds that the {\it rms} deviation for seven low-lying excitation states (the first two $0^+$, three $2^+$, and two $4^+$) is: $\sigma = 0.279$~MeV for the 5DCH, which is in comparison with the corresponding value 0.429 MeV obtained by using Nijmegen1 effective interaction in Ref.~\cite{Dikmen09}, and 0.137 MeV in  the calculation of Ref.~\cite{Guazzoni11}.

In the lower panel of Fig.~\ref{fig:BE2}, we present the $B(E2;
0^+_1\to 2^+_1)$ values from 5DCH calculations, in comparison with
the results by shell model and RQRPA, as well as the recently
measured data. It is shown that both the 5DCH and RQRPA calculations
reproduce rather well the systematics of $B(E2; 0^+_1\to 2^+_1)$
values in the whole isotopic chain, in particular the increase of
$E2$ strength from $^{116}$Sn down to $^{108}$Sn, which cannot be
described by the shell model. Quantitatively, the RQRPA slightly underestimates those $E2$ strengths around $^{116}$Sn.
While the 5DCH  overestimates  the $E2$ strength $\sim 0.1~e^2b^2$ when approaching to the neutron shell
closure, where the low-lying states are mainly caused by the excitation of few quasiparticles.
Concerning our interest and the power of our model, in the following we will mainly focus on the isotopes around middle shell,
where the excitation is dominated by the collective vibration and enhanced $E2$ transition is observed.

\begin{figure}[htb!]
\centering
\includegraphics[scale=0.28]{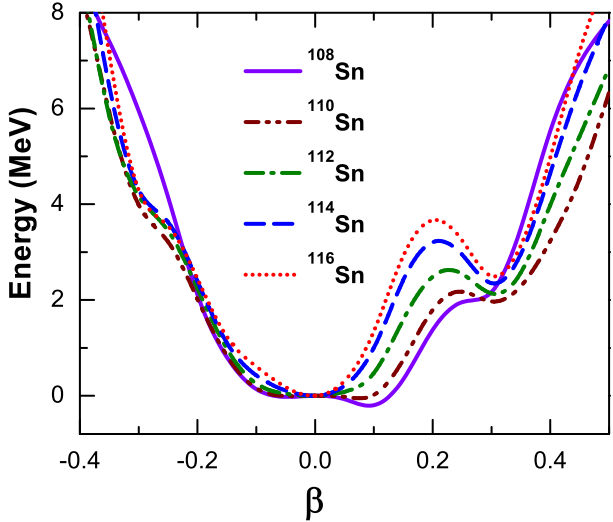}
\caption{\label{fig:PEC}(Color online) Self-consistent RMF+BCS
binding energy curves for the even-even $^{108-116}$Sn isotopes as
functions of the axial deformation parameter $\beta$. Energies are
normalized with respect to the binding energy of the corresponding
spherical state.}
\end{figure}

To understand the enhanced $E2$ transitions around $N=108$, we plot
the deformation energy curves of the even-even $^{108-116}$Sn
isotopes as functions of the axial deformation parameter $\beta$ in
Fig.~\ref{fig:PEC}. Due to the robust $Z=50$ proton shell, all the
Sn isotopes are spherical or near spherical. Around the neutron
mid-shell $N=66$, an excited prolate minimum with $\beta\sim 0.3$
shows up. As the decrease of the neutron number from
$N=66$ down to $N=58$, the energy curve around spherical shape
becomes softer gradually. In particular, for $^{108}$Sn, a weakly
prolate deformed global minimum with $\beta\sim0.1$ is observed.
In other words, the dynamic effect of quadrupole deformation fluctuations
is enlarging the nuclear collectivity gradually when decreasing neutron number
from $N=66$ down to $N=58$.

\begin{figure}[htb!]
\centering
\includegraphics[scale=0.26]{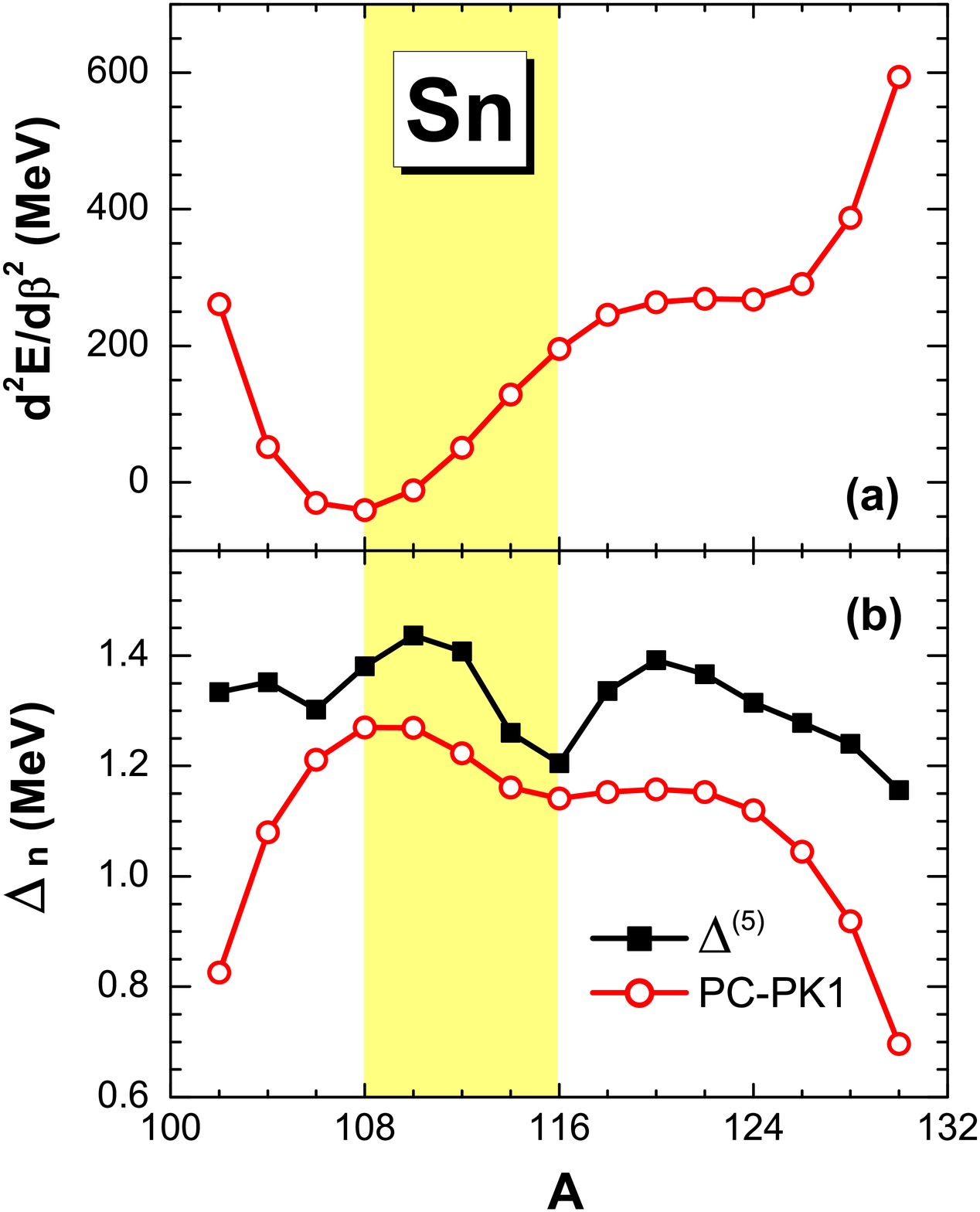}
\caption{\label{fig:CGP}(Color online) The predicted curvature of the
PEC (panel a) and average neutron pairing gaps (panel b) at spherical point
in Sn isotopic chain calculated by PC-PK1 density functional.
For comparison, the odd-even mass differences (denoted by solid squares) extracted
by the five-point formula are also plotted in panel b.}
\end{figure}

The upper panel of Fig.~\ref{fig:CGP} displays the curvatures of the
deformation energy curves around the spherical shape as a function
of neutron number. A rapid drop of the curvature in Sn isotopes with
$A<116$ is found, which is responsible for the increased $B(E2)$
value shown in Fig.~\ref{fig:BE2}. To understand this change in the
topograph of deformation energy curves, we plot the average neutron
pairing gaps, weighted by $uv$ coefficient as
Ref.~\cite{Bender02EPJA}, at spherical states in the lower panel of
Fig.~\ref{fig:CGP}. For comparison, the odd-even mass differences of
experimental data by the five-point formula are also plotted. It is
shown that the rapid drop of the curvature when decreasing neutron
number from $N=66$ is closely correlated to the neutron pairing
gaps. Similar evolution trend is shown in the odd-even mass
differences. It indicates that the strong pairings soften the energy
curve around spherical shape, which leads to the enhanced
collectivity in Sn isotopes around $N=58$.

\begin{figure}[htb!]
\centering
\includegraphics[scale=0.3]{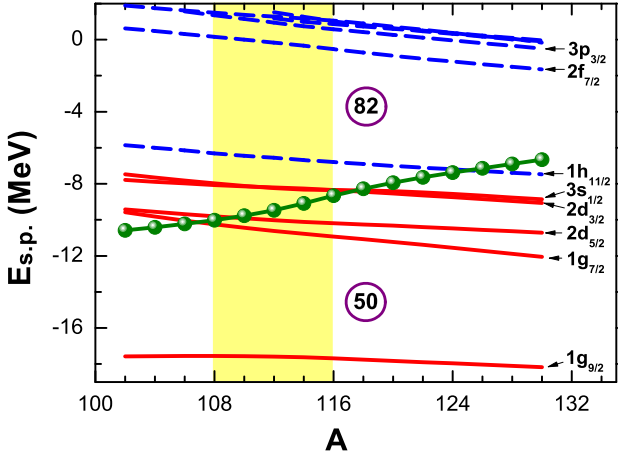}
\caption{\label{fig:sp} (Color online) Evolution of neutron single-particle levels
at spherical point for Sn isotopic chain. The balls indicate the position of Fermi
surface.}
\end{figure}

Figure~\ref{fig:sp} displays the neutron single-particle energies at
spherical shape along Sn isotopic chain. The balls indicate the
position of Fermi surface. It is seen that the levels $1g_{7/2}$ and
$2d_{5/2}$ move closer as the neutron number decreases from $N=82$
to $N=50$ shell. Meanwhile, the Fermi surface comes toward to these
two levels for the isotopes with $A<116$. This leads to the
increasing level density around Fermi surface and consequently
enhanced paring correlations. Moreover, it is found that the Fermi
surface crosses the $3s_{1/2}$ orbit in between $N=64$ and $N=70$.
The considerable occupation probability of $3s_{1/2}$ orbit hinders
the collectivity of these nuclei. Meanwhile, the obstructive effect
is further enhanced by the low density of single-particle states
around the Fermi surface, which reduces the pair correlations (c.f.
in Fig.~\ref{fig:CGP}). It provides a qualitative interpretation of
the dip of $E2$ transition at $^{116}$Sn, which is consistent with
that given in Ref. \cite{Jung.11}.

%
%

In summary, the low-lying collective states in Sn isotopes from
$N=50$ to $N=82$ have been studied by solving a five-dimensional
collective Hamiltonian with parameters determined from the triaxial
relativistic mean-field calculations. The calculations reproduce the
systematics of both excitation energies for $2^+_1$ states and
$B(E2;0^+_1\to 2^+_1)$  rather well. In particular, the enhancement
of $E2$ transitions from $N=66$ to $N=58$ has been found to be the
consequence of increasing quadrupole shape fluctuation
around spherical shape, induced by the  strengthening neutron pairing
correlations. This conclusion is consistent with that by the shell models,
but in the language of mean-field approach.
Furthermore, it is found the gradual degeneracy of levels $1g_{7/2}$ and
$2d_{5/2}$ around the Fermi surface that leads to the increase of
level density and consequently the enhanced paring correlations.
Moreover, it has shown that the Fermi surface crosses the $3s_{1/2}$
orbit in between $N=64$ and $N=70$, where the low level density and
considerable occupation probability of $3s_{1/2}$ orbit hinder the
nuclear collectivity and give rise to valley-like of the $E2$
transition as a function of neutron number.

%
\section*{Acknowledgments}

This work was supported in part by the Major State 973 Program 2013CB834400, the NSFC under Grant Nos. 10975008, 10947013, 11175002, 11105110, and 11105111, the Research Fund for the Doctoral Program of Higher Education under Grant No. 20110001110087, the Southwest University Initial Research Foundation Grant to Doctor (Nos. SWU110039, SWU109011), the Fundamental Research Funds for the Central Universities (XDJK2010B007 and DJK2011B002).

%
%

\bibliographystyle{model3-num-names}

\end{document}